\documentclass[12pt]{article}

\usepackage{graphics}
\usepackage{epsfig}
\usepackage{latexsym}
\usepackage{colordvi}
\usepackage{amsmath}
\usepackage{amssymb}

\def\simkl{^<\hskip -2.5mm_\sim}
\newcommand{\fun}{$F_2^{\gamma}(x,Q^2)$ }

\newcommand{\gam}{^{\gamma}}
\newcommand{\fund}{$F_2^{\gamma}(x,Q^2)$}
\newcommand{\be}{\begin{equation}}
\newcommand{\ba}{\begin{eqnarray}}
\newcommand{\ea}{\end{eqnarray}}
\newcommand{\etal}{{\it et al.}}

\newlength{\figwidth}
\newlength{\figheight}
\def\z0{Z}

\title{
\begin{flushright}
\begin{small}
IFT-2003-31
\end{small}
\end{flushright}
Uncertainties of the CJK 5 Flavour LO Parton Distributions in 
the Real Photon}
\author{
P.~Jankowski\\
{\small\it Institute of Theoretical Physics, Warsaw University, 
ul. Ho\.za 69, 00-681 Warsaw, Poland}\\
}
\date{}

\begin{document}
\maketitle

\begin{abstract}
Radiatively generated, LO quark ($u,d,s,c,b$) and gluon densities in the 
real, unpolarized photon, calculated in the CJK model being an improved 
realization of the CJKL approach, have been recently presented. The results 
were obtained through a global fit to the experimental \fun data. In this 
paper we present, obtained for the very first time in the photon case, an 
estimate of the uncertainties of the CJK parton distributions due to the 
experimental errors. The analysis is based on the Hessian method which was 
recently applied in the proton parton structure analysis. Sets of test 
parametrizations are given for the CJK model. They allow for calculation of 
its best fit parton distributions along with \fun and for computation of 
uncertainties of any physical value depending on the real photon parton 
densities. We test the applicability of the approach by comparing 
uncertainties of example cross-sections calculated in the Hessian and Lagrange
methods. Moreover, we present a detailed analysis of the $\chi^2$ of the CJK 
fit and its relation to the data. We show that large $\chi^2/_{\rm DOF}$ of 
the fit is due to only a few of the experimental measurements. By excluding 
them $\chi^2/_{\rm DOF}\approx 1$ can be obtained.
\end{abstract}

\clearpage


\section{Introduction}

In the recent paper, \cite{cjk}, new results on the LO unpolarized real 
photon parton distributions have been presented. In that work we improved and 
broadened our previous analysis, \cite{cjkl}. As a result three models and 
corresponding parametrizations of the photon structure function \fun and 
parton densities were given.

The main difference between the models lies in the way the heavy charm- and 
bottom-quark contributions to the photon structure function are described. In 
two approaches, referred to as FFNS$_{CJK}1$ and 2 we adopted a widely used 
scheme in which there are no heavy quarks, $h$, in the photon. In that case 
heavy quarks contribute to the photon structure function through the so-called 
Bethe-Heitler, $\gamma^* \gamma \to h\bar h$ interaction. The FFNS$_{CJK}2$ 
model includes an additional ``resolved'' contribution given by the 
$\gamma^* G \to h\bar h$ process, see for instance \cite{grst}. Finally, the 
model of our main interest, denoted as CJK, applies the ACOT$(\chi)$ 
\cite{acot} scheme, where heavy-quark densities appear.

The free parameters of each model have been computed by the means of the 
global fits to the set of updated \fun data collected in various $e^+e^-$ 
experiments. That way parametrizations of the photon structure function and
parton distributions were created.

The main goal of the present analysis is to estimate the uncertainties of the 
CJK parton distributions due to the experimental errors of \fun data. Alike in
the global CJK fit the raw experimental data are used, by which we mean that 
neither the radiative corrections, nor the corrections taking into the account
the small off-shelness of the probed quasi-real photons are included. Further,
the correlations among the measurements are neglected and in each $x$ bin the 
approximation $\langle F_2\gam(x)\rangle \approx F_2\gam(\langle x\rangle)$ is
assumed. Finally, one has to keep in mind that all the data were obtained with
an assumption of the FFNS scheme for the heavy-quark contributions to \fun. 
The inclusion of the above corrections is beyond the scope of this work.

Our work has been motivated by the recent analysis performed for the proton 
structure by the CTEQ Collaboration, \cite{cteq1}--\cite{cteq3} and the MRST 
group, \cite{mrst}. We use the Hessian method, formulated in recent papers, to
obtain sets of parton densities allowing along with the parton distributions 
of the best fit to calculate the best estimate and uncertainty of any 
observable depending on the photon structure. As the same aim can be obtained 
utilizing the Lagrange method, see for instance \cite{cteq1},\cite{mrst} and 
\cite{cteq4}, we apply this approach to obtain an independent test of the 
correctness of the Hessian model results.

Moreover we perform an analysis of the sources of the $\chi^2$ obtained in the
global fit for CJK model. Namely, we examine which $e^+e^-$ experiments data 
produce largest contributions to the total $\chi^2$ of this fit as well as for
other models including our FFNS$_{CJK}1,2$ approaches and GRS LO \cite{grs} 
and SaS1D \cite{sas} parametrizations.

The paper is divided into five parts. Section 2 shortly recalls the 
FFNS$_{CJK}$ and CJK models of the real photon structure. Section 3 is 
devoted to an introduction of the Hessian method and to a presentation of the 
calculated uncertainties of the CJK parton distributions. Furthermore, in 
section 4, we briefly introduce the Lagrange method. Next, in section 5, we 
compare the uncertainties of two example cross-sections calculated in both,
Hessian and Lagrange approaches. Finally, in section 6 we examine the data 
sources of the $\chi^2$ of the CJK fit. The parton distributions which are a 
result of our analysis have been parametrized on the grid. The FORTRAN 
programs obtained that way are open to the public use.


\section{FFNS$_{CJK}$ and CJK models - short recollection}

In this analysis we focus on the CJK model results. We are interested in the
parton distributions computed in this approach and most of all in their 
uncertainties due to the experimental data errors. Still, in order to estimate
the allowed deviation of the $\chi^2$ of the global CJK fit from its minimum
we will apply the FFNS$_{CJK}$ models. Moreover, they will be very useful in
determining of the $e^+e^-$-experiment data sets which are not well described
by the photon structure models. Therefore, we shortly recall the main 
differences between the FFNS type approaches and the CJK approach. Next, we 
present the origin of the free parameters of the CJK model calculated in 
\cite{cjk} by the means of the global fit to the \fun data. Further, the data 
used in the fits is described. Finally, we recall the results of the CJK model
fit.


\subsection{The models}

The two approaches leading to the FFNS$_{CJK}$ and CJK models, have been 
described in detail in our previous papers \cite{cjk} and \cite{cjkl}. The 
difference between them lays in the approach to the calculation of the heavy, 
charm- and beauty-quark contributions to the photon structure function \fund. 
First, FFNS$_{CJK}$ type models base on a widely adopted Fixed Flavour Number 
Scheme in which there are no heavy quarks in the photon. Their contributions 
to \fun are given by the 'direct' (Bethe-Heitler) 
$\gamma^* \gamma \to h\bar h$ process of heavy-quarks production $h$. In 
addition one can also include the so-called 'resolved'-photon contribution: 
$\gamma^* G \to h\bar h$. We denote these terms as 
$F_{2,h}\gam(x,Q^2)|_{direct}$ and $F_{2,h}\gam(x,Q^2)|_{resolved}$, 
respectively. We considered two FFNS models: in the first one, 
FFNS$_{CJK}$1, we neglected the resolved photon contribution, while in the 
second one, FFNS$_{CJK}$2, both mentioned contributions to \fun were 
included. In such models the heavy-quark masses are kept to their physical 
values. The photon structure function is then computed as
\be
\label{F2FFNS}
F_2\gam(x,Q^2) = \sum_{i=1}^3 xe_i^2 (q_i\gam +\bar q_i\gam)(x,Q^2) +
\sum_{h(=c,b)} \left[ F_{2,h}^{\gamma}(x,Q^2)|_{direct} + 
F_{2,h}\gam(x,Q^2)|_{resolved} \right],
\end{equation}
with $q\gam_i(x,Q^2)$ ($\bar q\gam_i(x,Q^2)$) being the light $u,d,s$ quark 
(anti-quark) densities, governed by the 
Dokshitzer-Gribov-Lipatov-Altarelli-Parisi (DGLAP) evolution equations.

The CJK model adopts the new ACOT($\chi$) scheme, \cite{acot}, which is a 
recent realization of the Variable Flavour Number Scheme (VFNS). In this scheme
one combines the Zero Mass Variable Flavour Number Scheme (ZVFNS), where
the heavy quarks are considered as massless partons of the photon, with
the FFNS just discussed above. In this model, in addition to the terms shown 
in Eq. (\ref{F2FFNS}) one must include the contributions due to the 
heavy-quark densities which now appear also in the DGLAP evolution equations.
A double counting of the heavy-quark contributions to \fun must be corrected 
with the introduction of subtraction terms for both, the direct and 
resolved-photon, contributions. Further, following the ACOT($\chi$) scheme, we
introduced the $\chi_h=x(1+4m_h^2/Q^2)$ parameters responsible for the proper 
vanishing of the heavy-quark densities at the kinematic thresholds for their 
production in DIS: \\ $W^2=Q^2(1-x)/x>4m_h^2$, where $W$ is the 
$\gamma^* \gamma$ 
centre of mass energy. Adequate substitution of $x$ with $\chi_h$ in $q_h$ and
the subtraction terms forces their correct threshold behavior as 
$\chi_h \to 1$ when $W \to 2m_h$. This is achieved for all the terms except 
for the subtraction term $F_{2,h}^{\gamma}(x,Q^2)|_{dir,subtr}$ which
requires an additional constraint, $F_{2,h}^{\gamma}(x,Q^2)|_{dir,subtr}=0$ 
for $\chi_h>1$, imposed by hand. Finally, we obtain the following formula 
for the photon structure function in the CJK model
\ba
F_2^{\gamma}(x,Q^2) &=& x\sum_{i=1}^3 e_i^2 (q_i\gam +\bar q_i\gam)(x,Q^2) +
x\sum_{h(=c,b)} e_h^2 (q_h\gam + \bar q_h\gam)(\chi_h,Q^2) \nonumber \\
&& + \sum_{h(=c,b)} \left[ F_{2,h}^{\gamma}(x,Q^2)|_{direct} + 
F_{2,h}^{\gamma}(x,Q^2)|_{resolved} \right] \\
&& - \sum_{h(=c,b)} \left[ F^\gamma_{2,h}|_{dir,subtr} (\chi_h,Q^2)
      + F^\gamma_{2,h}|_{res,subtr} (\chi_h,Q^2) \right]. \nonumber
\label{finalacot}
\ea

The above formula for \fun must be complited by imposing of another condition 
(the positivity constraint):
\begin{equation} 
F_{2,h}\gam(x,Q^2) \geq F_{2,h}\gam(x,Q^2)|_{direct}
                      + F_{2,h}\gam(x,Q^2)|_{resolved}
\end{equation}
preventing the unphysical situation $F_{2,h}\gam(x,Q^2)<F_{2,h}\gam(x,Q^2)|_{direct}+F_{2,h}\gam(x,Q^2)|_{resolved}$ which can occur at small- and large-$x$.

Explicit expressions for the terms appearing in Eqs. (\ref{F2FFNS}) and 
(\ref{finalacot}) can be found in \cite{cjkl}.

For all models we chose to start the DGLAP evolution at a small value 
of the $Q^2$ scale, $Q_0^2=0.25$ GeV$^2$, hence our parton densities are 
radiatively generated. As it is well known the point-like contributions 
are calculable without further assumptions, while the hadronic parts need the 
input distributions. For that purpose we utilize the Vector Meson Dominance 
(VMD) model \cite{VMD}, with the assumption that all light vector meson 
contributions are proportional to the $\rho^0$ meson contribution and are
accounted for via a parameter $\kappa$, that is left as a free parameter. We 
took parton densities in the photon equal to
\begin{equation}
f_{had}\gam(x,Q_0^2) = 
\kappa\frac{4\pi \alpha}{\hat f^2_{\rho}}f^{\rho}(x,Q_0^2).
\label{vmdfor}
\end{equation}
The parameter $\hat f^2_{\rho}$ is extracted from the experimental data on the
$\Gamma(\rho^0 \to e^+e^-)$ width.

We assumed the input densities of the $\rho^0$ meson at $Q_0^2=0.25$ GeV$^2$ in
the form of valence-like distributions both for the (light) quark 
($v^{\rho}$) and gluon ($G^{\rho}$) densities. All sea-quark distributions 
(denoted as $\zeta^{\rho}$), including $s$-quarks, were neglected at the input 
scale:
\ba
xv^{\rho}(x,Q_0^2) &=& N_v x^{\alpha}(1-x)^{\beta},  \\
xG^{\rho}(x,Q_0^2) &=& \tilde N_g xv^{\rho}(x,Q_0^2)= 
N_g x^{\alpha}(1-x)^{\beta}, \\
x \zeta^{\rho}(x,Q_0^2)&=&0, 
\label{input1}
\ea
where $N_g=\tilde N_gN_v$.

The valence-quark and gluon densities must satisfy the constraint representing 
the energy-momentum sum rule for $\rho$:
\be
\int_0^1 x(2v^{\rho}(x,Q_0^2)+G^{\rho}(x,Q_0^2))dx = 1.
\label{const2}
\end{equation}
That constraint allowed to express in the case of the CJK model the 
normalization factor $N_g$ as a function of $\alpha, \beta, N_v$ and $\kappa$.

In this analysis we decided to relax the second possible constraint, 
related to the number of valence quarks in the $\rho^0$ meson, $n_v$:
\be
n_v = \int_0^1 2v^{\rho}(x,Q_0^2)dx = 2.
\end{equation}
We observed that imposing that constraint spoils the quadratic approximation
of the Hessian method described in section \ref{secthess}. The reason may be 
the additional correlation between the $\alpha$ and $\beta$ parameters which
appears when we express the $N_v$ parameter as a function of the above two.
On the other hand, in the CJK model case, the global fit without imposing
the additional constraint gives proper value, $n_v=2 \pm 0.1$. Therefore we
decided to use it as the CJK fit and further apply the Hessian method to 
examine the uncertainties of the resulting parton distributions.


\subsection{Data \label{datasect}}

The CJK as well as the FFNS$_{CJK}$ model fits were performed using all 
existing \fun data, \cite{CELLO}--\cite{HQ2}, apart from the old TPC2$\gamma$,
\cite{TPC}. In our former global analysis \cite{cjkl} we used 208 \fun 
experimental points. Now we decided to exclude the TPC2$\gamma$ data from the 
set because it has been pointed out that these data are not in agreement with 
other measurements (see for instance \cite{klasen}). After the exclusion we 
are left with 182 \fun experimental points.

The raw experimental data are used, by which we mean that neither the radiative
corrections, nor the corrections taking into the account the small 
off-shelness of the quasi-real photons probed in the experiments are included. 
Further, the correlations among the measurements, not presented in most of the
articles \cite{CELLO}--\cite{HQ2}, are neglected. Finally, in each $x$ bin the
approximation $\langle F_2\gam(x)\rangle \approx F_2\gam(\langle x\rangle)$ is
assumed.

We included all the data in the $\chi^2$ fit without any weights. A list of 
all experimental points used can be found on the web-page \cite{webpage}.


\subsection{Results of the CJK global fit}

The fit of the CJK model to the experimental \fun data set described above was
based on the least-squares principle (minimum of $\chi^2$) and were done using 
\textsc{Minuit} \cite{minuit}. Systematic and statistical errors on data 
points were added in quadrature.

The results of our new fit are presented in table \ref{tparam}. The second 
and third columns show the quality of the fit, i.e. the total $\chi^2$ for
182 points and the $\chi^2$ per degree of freedom. The fitted values for
parameters $\alpha$, $\beta$, $\kappa$ and $N_v$ are presented in the middle
of the table with the errors obtained from \textsc{Minos} with the standard 
requirement of $\Delta \chi^2 = 1$. In addition, the value for $\tilde N_g$ 
obtained from other parameters using the constraint (\ref{const2}) is given in
the last column.

\begin{table}[htb]
\begin{center}
\renewcommand{\arraystretch}{1.5}
\begin{tabular}{|c|@{} p{0.05cm} @{}|c|c|@{} p{0.08cm} @{}|c|c|c|c|@{} p{0.08cm} @{}|c|}
\hline
 model           && $\chi^2$ (182 pts) & $\chi^2/_{\rm DOF}$ && $\kappa$ & $\alpha$ & $\beta$ & $N_v$ && $ \tilde N_g$ \\
\hline
\hline
 CJK           &&        273.7       &         1.537       &&  $1.934^{+0.131}_{-0.124}$  & $0.299^{+0.077}_{-0.069}$ & $0.898^{+0.316}_{-0.275}$ & $0.404^{+0.116}_{-0.088}$ && 4.93  \\
\hline
\end{tabular}
\caption{The total $\chi^2$ for 182 data points used in the fit and for the 
degree of freedom and parameters of the fit for the CJK model. All the given 
errors are obtained from \textsc{Minos} with the standard requirement of 
$\Delta \chi^2 = 1$.}
\label{tparam}
\end{center}
\end{table}

The results of the CJK fit, its agreement with various experimental data 
and its comparison to the other models and other photon structure 
parametrizations have been described in detail in \cite{cjk}. In this paper we
focus on the uncertainties of the parton distributions due to the experimental
\fun data. Moreover, we check in detail the influence of various measurements 
on the quality of the fit.


\section{Uncertainties of the parton distributions - Hessian method 
\label{secthess}}

We are going to address in this article a problem that so far has never
been considered in the case of the photon structure. Namely, we want to 
present an analysis of the experimental uncertainties of our CJK parton 
densities. To reach that goal we derive the necessary knowledge and tools from
the proton structure considerations.

During the last two years numerous analysis of the uncertainties of the proton
parton densities resulting from the experimental data errors appeared. The 
CTEQ Collaboration in a series of publications, \cite{cteq1}-\cite{cteq3}, 
developed and applied a new method of their treatment significantly improving 
the traditional approach to this matter. Later the same formalism has been 
applied by the MRST group in \cite{mrst}. The method bases on the Hessian 
formalism and as a result one obtains a set of parametrizations allowing for 
the calculation of the uncertainty of any physical observable depending on the 
parton densities.

The Hessian method is a very useful tool as it allows for computing of the
parton density uncertainties in a very simple and effective way. Still, it
relies on the assumption of the quadratic approximation which not necessary is
perfectly preserved.


\subsection{The method}

A detailed description of the method applied in our analysis can be found
in \cite{cteq1} and \cite{cteq2}. For the sake of clearness of our procedures
we will partly repeat it here, keeping the notation introduced by the 
CTEQ Collaboration.

Let us consider a global fit to the experimental data based on the 
least-squares principle performed in a model, being parametrized with a set of 
$\{a_i,i=1,2,\cdots,d\}$ parameters. Each set of values of these parameters 
constitutes a test parametrization $S$. The set of the best values of 
parameters $\{a_i^0\}$, corresponding to the minimal $\chi^2$, $\chi_0^2$,
is denoted as $S^0$ parametrization. In the Hessian method one makes a basic 
assumption that the deviation of the global fit from $\chi_0^2$ can be 
approximated in its proximity by a quadratic expansion in the basis of
parameters $\{a_i\}$
\be
\Delta \chi^2 = \chi^2-\chi_0^2 = \sum_{i=1}^d \sum_{j=1}^d H_{ij}(a_i-a_i^0)
(a_j-a_j^0),
\end{equation}
where $H_{ij}$ is an element of the Hessian matrix calculated as
\be
H_{ij} = \frac{1}{2}\left( \frac{\partial^2\chi^2}{\partial a_i \partial a_j} 
\right)_{a_0}.
\label{hess}
\end{equation}

Since the $H_{ij}$ is a symmetric matrix, it has a complete set of 
$k=1,2,\cdots,d$ orthonormal eigenvectors $(v_i)_k$ defined by
\ba
\sum_{j=1}^d H_{ij}(v_j)_k &=& \epsilon_k (v_i)_k, \\
\sum_{i=1}^d (v_i)_j(v_i)_k &=& \delta_{jk},
\ea
with $\{\epsilon_k\}$ being the corresponding eigenvalues. Variations around
the minimum can be expressed in terms of the basis provided by the set of
eigenvectors
\be
a_i-a_i^0 = \sum_{k=1}^d s_k z_k (v_i)_k ,
\label{twobasis}
\end{equation}
where $\{z_k\}$ are new parameters describing the displacement from the best 
fit. The $\{s_k\}$ are scale factors introduced to normalize $\{z_k\}$ in 
such a way that
\be
\Delta \chi^2 = \chi^2-\chi_0^2 = \sum_{k=1}^d z_k^2.
\label{quadrz}
\end{equation}
The above equation means that the surfaces of constant $\chi^2$ are spheres 
in $\{z_k\}$ space. That way the $\{z_k\}$ coordinates create a very useful, 
tenormalized basis. The $(v_i)_k \equiv v_{ik}$ matrix describes the 
transformation between this new basis $\{z_k\}$ and the old $\{a_i\}$ 
parameter basis. The scaling factors $s_k$ are equal to $\sqrt{1/\epsilon_k}$ 
provided that we work in the ideal quadratic approximation. In reality they 
differ from these values.

The Hessian matrix can be calculated from its definition in Eq. (\ref{hess}). 
Such a computation meets many practical problems arising from the large range 
spanned by the eigenvalues $\{\epsilon_k\}$, the numerical noise and 
non-quadratic contributions to $\chi^2$. The solution has been given by 
the CTEQ Collaboration \cite{cteq1} which introduced an iterative procedure 
working properly in the presence of all the problems listed above. The 
algorithm has been implemented as an extension to the \textsc{Minuit} program 
and is open for public use (see \cite{cteq1}).

Having calculated the eigenvectors, eigenvalues and scaling factors we can
create a basis of the parametrizations of the parton densities, 
$\{S_k^{\pm},k=1,\cdots,d\}$. Each of the $S_k^{\pm}$ pairs corresponds
to a different eigenvector direction $(v_i)_k$. These parametrizations are 
constructed in the following way: their parameters are defined by 
displacements of a magnitude $t$ ``up'' or ``down'' along the corresponding 
eigenvector direction
\be
a_i(S_k^{\pm}) = a_i^0 \pm t \: (v_i)_k s_k.
\label{delpar}
\end{equation}
For example the set of $S_1^+$ parameters is given by inserting
$(z_1,z_2,z_3,\cdots)=(t,0,0,\cdots)$ into Eq. (\ref{twobasis}). For each 
$S_k^{\pm}$ parametrization $\Delta \chi^2 = t^2$.

Further, let us consider a physical observable $X$ depending on the photon 
parton distributions. Its best value is given as $X(S^0)$. The uncertainty
of $X$, for a displacement from the parton densities minimum by 
$\Delta \chi^2 = T^2$ (T - the tolerance parameter) can be calculated with a 
very simple expression (named as master equation by the CTEQ Collaboration)
\be
\Delta X = \frac{T}{2t}\left( \sum_{k=1}^d[X(S_k^+)-X(S_k^-)]^2 \right)
^{\frac{1}{2}}.
\label{master}
\end{equation}
Note that having calculated $\Delta X$ for one value of the tolerance parameter
$T$ we can obtain the uncertainty of $X$ for any other $T$ by simple scaling 
of $\Delta X$. This way sets of $\{S_k^{\pm}\}$ parton densities give us a 
perfect tool for studying of the uncertainties of other physical quantities. 
One of such quantities can be the parton densities themselves.

Finally, we can calculate the uncertainties of the $a_i$ parameters of the 
model. According to Eq. (\ref{delpar}) in this case 
$a_i(S_k^+)-a_i(S_k^-) = 2t (v_i)_k s_k$ and the master equation gives a 
simple expression
\be
\Delta a_i = T\left( \sum_{k=1}^d v_{ik} s_k \right)^{\frac{1}{2}}.
\label{maspar}
\end{equation}

We end this shortened description of the used method with one practical point. 
In real analysis we observe the considerable deviations from the ideal 
quadratic approximation of equation (\ref{quadrz}). To make an improvement we 
can adjust the scaling factors $\{s_k\}$ either to obtain exactly 
$\Delta \chi^2 = t^2$ at $z_k=t$ for each of the $S_k^{\pm}$ sets or to get 
the best average agreement over some $z_k$ range (for instance for 
$z_k \le t$). We chose to apply the second approach.


\subsection{Estimate of the tolerance parameter $T$ for the CJK photon 
densities \label{secttolpar}}

We consider now the value of the tolerance parameter $T$ defining the allowed
deviation of the global fit from the minimum, $\Delta \chi^2 = T^2$, as 
described in the previous section. Through the master equation (\ref{maspar}) 
$T$ is relevant to the calculations of the uncertainties due to the real 
photon parton densities. In case of an ideal analysis $\Delta \chi^2=1$ is a 
standard requirement. Of course a global fit to the \fun data coming from 
various experiments is not such a case and certainly $T$ must be greater 
than 1. Unfortunately no strict rules allowing for estimation of the tolerance
parameter exist. A detail analysis of that problem can be found in 
\cite{cteq2} and \cite{cteq4}. We try to estimate the reasonable practical $T$
value for the CJK fit in two ways.
sa

First we examine the mutual compatibility of the experiments used in the fit. 
We divide the data into six sets. Four of them contain the results of the 
CERN-LEP accelerator experiments - ALEPH, DELPHI, L3 and OPAL. The other two 
are sets of data collected with the DESY-PETRA and KEK-TRISTAN accelerators.
The DESY-PETRA collection, referred to as DESY, combines the results of PLUTO, 
JADE, CELLO and TASSO collaborations. The KEK-TRISTAN set, denoted as KEK, 
combines the results of TOPAZ and AMY experiments. For each of the data 
sets we calculate the $\chi^2_n$ and $\chi^2_{-n}$ (for the $n$th collection),
values of the $\chi^2$ of the best fit corresponding to the given experiment 
and to the remaining five ones respectively. Further we test how much 
$\chi^2_{-n}$ can be lowered by minimazing $\chi^2$ with the removed $n$th 
set of data. We obtain $\Delta \chi^2_{-n}$ value which is the minimal 
deviation of the global fit from its minimum necessary to describe the 
inclusion of the $n$th experiment to the global set of data. Results of that 
test are presented in table \ref{chi2t}. The $\chi^2_n/N_n$ values, where
$N_n$ is a number of the experimental points in the $n$-th data set, indicate 
that truly our global fit is not a case of an ideal analysis. For some of 
experiments our CJK fit agrees very well with the data. In case of 
others $\chi^2_n/N_n$ is much larger than 1 and $\chi^2/_{\rm DOF} = 1.537$ of 
the global CJK fit presented in table \ref{tparam}. Further we see that the 
$\Delta \chi^2_{-n}$ varies from 0.1 to 20.9 obtaining the maximal value in 
the DELPHI experiments case. For full reliability of the test we performed an 
additional fit including all available \fun data which means that apart from 
all the experiments used in the CJK fit and mentioned above the TPC2$\gamma$ 
data was utilized. We checked then that the $\Delta \chi^2_{-n}$ for the 
TPC2$\gamma$ case equals 3.2. We notice that the $\Delta \chi^2_{-n}$ values 
in the case of DESY, DELPHI and OPAL sets are larger. One of the reasons for 
that may be the following: the $F_2\gam(x,Q^2)$ points measured by the 
TPC2$\gamma$ experiment lie mostly in the low-$Q^2$ region not covered by 
other measurements. That is not the case for the other experiments and so if 
their results differ the exclusion of one of them can strongly effect the 
$\chi^2$ of the fit. Anyway, though there may exist other arguments for that,
it seems difficult to support the statement that the TPC2$\gamma$ data are 
inconsistent with other measurements as claimed in \cite{klasen}. Still, please
notice that as described in section \ref{datasect} the raw experimental data
were applied and their various possible corrections as well as the 
correlations among the measurements were not taken into consideration in our 
analysis. More discussion on the data will be given in the section 
\ref{sectdata}. Finally, we have to assume that the allowed $\Delta \chi^2$ is
greater than 20.9 and hence the tolerance parameter must have value $T \sim 5$.

\begin{table}[htb]
\begin{center}
\begin{tabular}{|c|c|@{} p{0.1cm} @{}|c|c|c|c|c|c|} 
\hline
n & Set    && \# of points $N_n$ & $\chi^2_n$ &$\chi^2_n/N_n$ & $\chi^2_{-n}$ & $\Delta \chi^2_{-n}$ \\
\hline
\hline
1 & DESY   && 38 & 89.5 & 2.36 & 184.2 & 9.6  \\
\hline
2 & KEK    && 16 & 18.2 & 1.14 & 255.5 & 0.2  \\
\hline
3 & ALEPH  && 20 & 21.0 & 1.05 & 252.7 & 1.3  \\
\hline
4 & DELPHI && 38 & 88.6 & 2.33 & 185.1 & 20.9 \\
\hline
5 & L3     && 28 & 14.9 & 0.53 & 258.8 & 0.6  \\
\hline
6 & OPAL   && 42 & 41.3 & 0.98 & 232.4 & 5.8  \\
\hline
\hline 
\end{tabular}
\caption{The CJK model. Table presents number of the points, $N_n$ in each 
data set, $\chi^2_n$ (and $\chi^2_n/N_n$) and $\chi^2_{-n}$, being values of 
the $\chi^2$ of the best fit corresponding to the $n$th set and to the 
remaining five sets respectively. Finally $\Delta \chi^2_n$ is the value by 
which $\chi^2_{-n}$ can be improved by the additional minimization.}
\label{chi2t}
\end{center}
\end{table}

As a second test we compare the results of our three fits presented in this 
paper. We find the $T$ values for which parton densities predicted by the FFNS
models lie between the lines of uncertainties of the CJK model parton 
distributions. We made such test independently for each of the five quark 
flavours and for the gluon densities in three $Q^2$ ranges and for 
$10^{-5} \le x \le 0.97$. As can be seen in table \ref{denst} the resulting 
numbers differ for various parton densities. They also depend on the $Q^2$ 
range in which we studied the $T$ parameter. While not strong in the case of 
quark distributions this dependence is dramatic in the gluon case predicting 
large $T$ value necessary to contain other gluon distributions within the CJK 
uncertainty-bands at high $Q^2$.

We do not think that the comparison of the \fun values predicted by various 
models and parametrizations should give us better information on the necessary 
$T$ parameter than the above test done for the parton distributions case.
\fun predictions seem to be much more model dependent than the resulting 
parton distributions. As an example we can mention the difference between the 
FFNS$_{CJK}$ fits ($\Delta \chi^2 \approx 34$ - see table \ref{denst}) which 
is not so apparent in the differences between the corresponding parton 
distributions. That can be easily understood when we recall that the 
FFNS$_{CJK}$ 1 and 2 models differ only by the resolved 
$\gamma^* G\to h\bar h$ term which contributes to the \fun but not to the 
DGLAP equations and therefore directly does not influence the parton densities.

\begin{table}[htb]
\begin{center}
\begin{tabular}{|c|@{} p{0.1cm} @{}|c|c|c|c|c|}
\hline
  && T($G\gam$) & T($d\gam$) & T($u\gam$) & T($s\gam$) & T($F_2\gam$) \\
\hline
\hline
$1 \le Q^2 \le 100$ GeV$^2$     &&  4.5  & 7.0 & 7.0 & 3.4 & 8 \\
\hline
$1 \le Q^2 \le 1000$ GeV$^2$    && 14.0  & 7.0 & 7.0 & 3.4 & 10.5 \\
\hline
$1 \le Q^2 \le 200000$ GeV$^2$  && 138.0 & 7.0 & 7.0 & 3.4 & 20.0 \\
\hline
\hline
\end{tabular}
\caption{The $T(q\gam)$ values for which parton $q\gam(x,Q^2)$ densities 
predicted by the FFNS models lie between the lines of uncertainties 
of the CJK model parton distributions. Calculation performed for various 
$Q^2$ ranges and for $10^{-5} \le x \le 0.97$.}
\label{denst}
\end{center}
\end{table}

We estimate that the tolerance parameter $T$ should lie in the range 
$5\sim 10$. Still, one has to keep in mind that because of the lack of data 
constraining the real photon gluon distribution, in case of processes 
dominated by the gluon interactions such assumption may not be safe enough.


\subsection{Tests of quadratic approximation}

As a result of application of the Hessian method to the CJK model we obtained 
a set of the $\{(v_i)_k\}$ and $\{s_k\}$ values with $i$ and $k=1,\cdots 4$ 
which corresponds to the number of free parameters ($\kappa$, $\alpha$, 
$\beta$ and $N_v$) in the model. We used the iteration procedure provided by 
the CTEQ Collaboration (see \cite{cteq1}). The procedure was run with 
displacements giving $\Delta \chi^2=1,3,5$ and 10. Each time 15 iterations 
were performed. Tests of the quadratic approximation described in detail below 
(for the case of final results) allowed us to choose the results obtained with
the $\Delta \chi^2 = 5$ displacement as the most reliable. Further we adjusted
the scaling factors $\{s_k\}$ to improve the average quadratic approximation 
over the $z_k \le 5$ range. The necessary additional factors multiplying the
$s_k$ values are shown in table \ref{mult}. We see that apart from the numbers
corresponding to the last eigenvector direction no significant adjustment is 
needed.

\begin{table}[htb]
\begin{center}
\begin{tabular}{|c|c|c|c|c|c|}
\hline
 model  &  direction  & \multicolumn{4}{c|}{eigenvector} \\
\cline{3-6}
        &  ``up'' or ``down'' & 1 & 2 & 3 & 4 \\
\hline
\hline
 CJK    & $S^+$       & 1.04   &  1.02  &  1.03 &  0.96  \\
\cline{2-6}
        & $S^-$       & 0.96   &  0.98  &  0.98 &  0.73  \\
\hline
\hline
\end{tabular}
\caption{Multiplication factors improving the average 
quadratic approximation in the $z_k \le 5$ range corresponding to all 
eigenvector $(v_i)_k$ directions and for all $\{S_k^{\pm}\}$ parametrizations
for the CJK model.}
\label{mult}
\end{center}
\end{table}

To be certain that our results are correct we need to check if the 
quadratic approximation on which the Hessian method relies is valid in the 
considered $\Delta \chi^2$ range.

\begin{figure}[htb]
\includegraphics[scale=1.0]{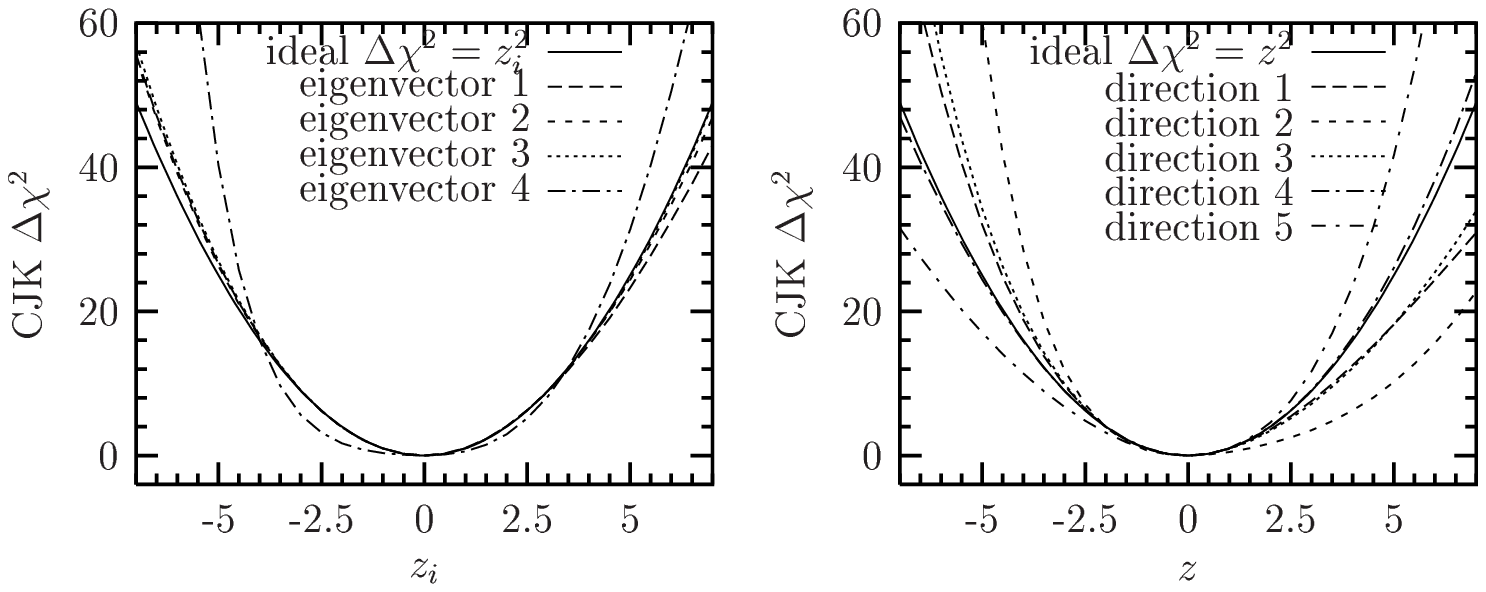}%
\vskip -0.7cm
\caption{Left plot presents comparison of the $\chi^2$ dependence
along each of four eigenvector directions to the ideal $\Delta \chi^2 = z_i^2$ 
curve. In right plot analogous comparison for 5 random directions
in $\{z_i\}$ space are shown. The ideal curve corresponds to 
$\Delta \chi^2 = \sum z_i^2 = z^2$}
\label{quad11}
\end{figure}

\begin{figure}[htb]
\includegraphics[scale=1.0]{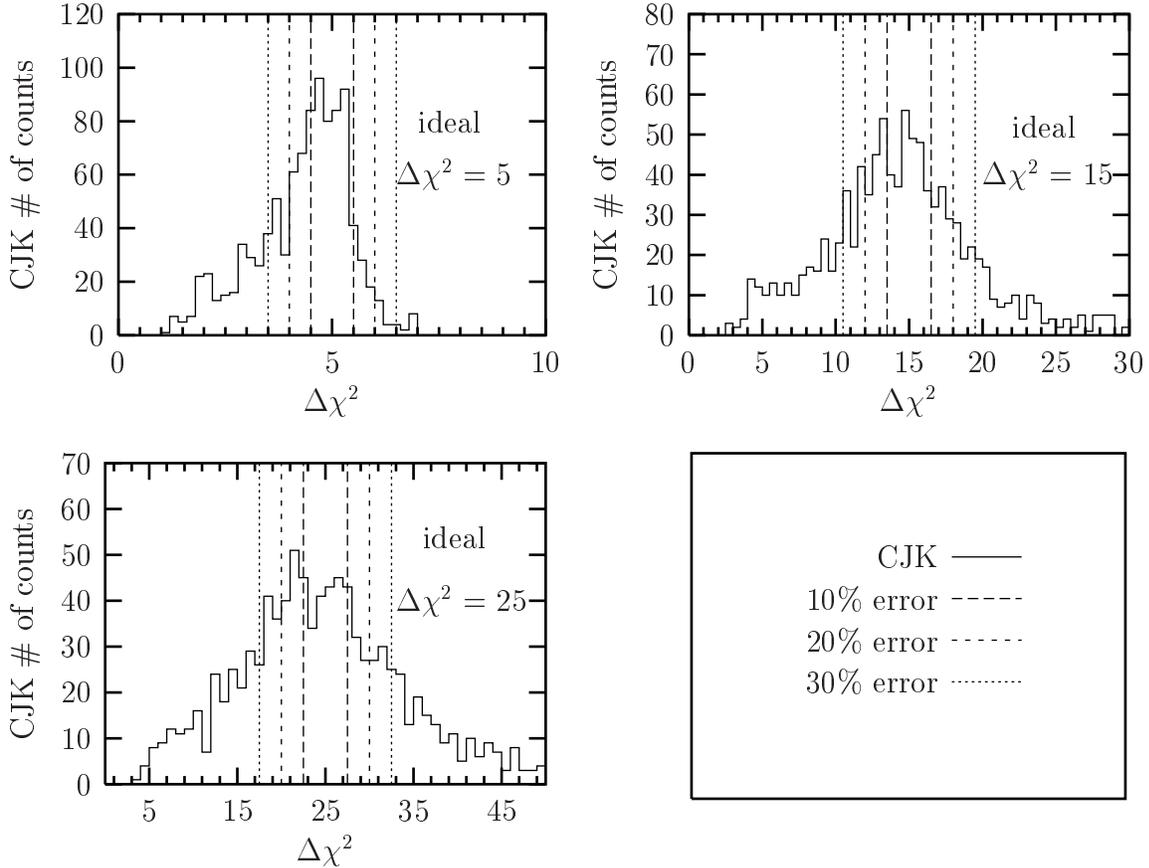}%
\vskip -0.7cm
\caption{Test of the frequency distribution of $\Delta \chi^2$ for 1000 
randomly chosen directions in the ${z_i}$ space normalized so that they 
correspond to the ideal $\Delta \chi^2 = 5, 15$ or 25. The 10, 20 and 30\% 
deviations from the expected $\Delta \chi^2$ are indicated in each plot.}
\label{quad12}
\end{figure}

In the left plot of Fig. \ref{quad11} we present the comparison of the 
$\chi^2$ dependence along each of four eigenvector directions 
($z_i=\delta_{ik}$ for the eigenvector $k$) to the ideal 
$\Delta \chi^2 = z_i^2$ curve. We see again that only the line corresponding 
to the eigenvector 4 does not agree with the theoretical prediction. Moreover 
it has a different shape than other lines which results from the scaling 
adjustment procedure. In the right plot of Fig. \ref{quad11} analogous 
comparison for the 5 randomly chosen directions in the $\{z_i\}$ space is 
shown. For each of directions $\sum_{k=1}^4 z_k^2=z^2$ and the ideal curve 
corresponds to $\Delta \chi^2 = z^2$. In this case we observe greater 
deviation from the quadratic approximation. Finally we test the frequency 
distribution of $\Delta \chi^2$ for randomly chosen directions in the 
$\{z_i\}$ space. Figure \ref{quad12} presents the frequency distribution for 
1000 $\{z_i\}$ directions normalized in such a way that they correspond to the
ideal $\Delta \chi^2 = 5, 15$ or 25 ($z=\sqrt{5},\sqrt{15}$ or 5 
respectively). The 10, 20 and 30\% deviations from the theoretical 
$\Delta \chi^2$ are indicated in each plot. We observe that, as could be 
expected, the quadratic approximation worsens with increasing $z$. Still the 
central peak is well outlined even in the last histogram. Fractions of the 
counts within the successive lines of the increasing deviation from the 
theoretical predictions are given in table \ref{freq}. Even for high 
$\Delta \chi^2 = 25$ more than half of the counts are contained in the 30\% 
error range.

\begin{table}[htb]
\begin{center}
\begin{tabular}{|c|c|c|c|}
\hline
                      & 10\%  & 20\%  & 30\% \\
\hline
\hline
 $\Delta\chi^2 = 5$   & 0.477 & 0.652 & 0.792   \\
\hline
 $\Delta\chi^2 = 15$  & 0.266 & 0.498 & 0.667   \\
\hline
 $\Delta\chi^2 = 25$  & 0.251 & 0.401 & 0.586   \\
\hline
\hline
\end{tabular}
\caption{Fractions of the counts within the successive lines of the 
increasing deviation from the ideal $\Delta \chi^2$ for 1000 randomly chosen 
directions in the $\{z_i\}$ space (see text and figure \ref{quad12}) for
the CJK model.}
\label{freq}
\end{center}
\end{table}


\subsection{Collection of test CJK parametrizations}

After checking that the Hessian method gives reasonable results for the case
of our CJK model we can create a collection of test parametrizations of the 
parton densities, $\{S_i^{\pm}\}$, for that model. We only need to choose the 
value of the magnitude $t$ of equations (\ref{delpar}) and (\ref{master}). We 
performed tests with $t=1,\sqrt{5},\sqrt{10}$ and 5 and noticed hardly any 
dependence of the uncertainties of parton densities or example physical 
cross-sections on its choice. We decided to apply $t=5$. That way the 
$\{S_i^{\pm}\}$ collection of eight parametrizations containing information 
about the CJK fit uncertainties was created. All sets of parameters are 
presented in table \ref{cjkpar}. 

\begin{table}[htb]
\begin{center}
\begin{tabular}{|c|c|@{} p{0.1cm} @{}|c|c|c|c|}
\hline
 eigenvector   & Set    && $\kappa$ & $\alpha$ & $\beta$ & $N_v$ \\
\hline
\hline
 1             & Set $1^+$ && 1.916    &  0.344   &  0.893  &  0.377 \\
\cline{2-7}
               & Set $1^-$ && 1.952    &  0.258   &  0.903  &  0.430 \\
\hline
 2             & Set $2^+$ && 1.937    &  0.378   &  0.841  &  0.544 \\
\cline{2-7}
               & Set $2^-$ && 1.932    &  0.224   &  0.952  &  0.271 \\
\hline
 3             & Set $3^+$ && 2.462    &  0.480   &  0.975  &  0.323 \\
\cline{2-7}
               & Set $3^-$ && 1.437    &  0.129   &  0.826  &  0.481 \\
\hline
 4             & Set $4^+$ && 1.754    &  0.525   &  1.912  &  0.691 \\
\cline{2-7}
               & Set $4^-$ && 2.072    &  0.127   &  0.124  &  0.185 \\
\hline
\end{tabular}
\caption{Parameters of the collection of CJK parametrizations.}
\label{cjkpar}
\end{center}
\end{table}

Next, in table \ref{parerrors} we present again the parameters of the CJK 
model with their errors calculated within the Hessian quadratic approximation
for the standard requirement of $\Delta \chi^2 = 1$. These uncertainties
should be compared with the errors calculated by \textsc{Minuit} and shown in 
table \ref{tparam}. They are of the same order but slightly smaller. Note that
the uncertainties given in table \ref{parerrors} can be simply multiplied by 
the tolerance parameter $T$ to obtain errors which would correspond to an 
assumption of a higher $\Delta \chi^2$ value (see Eq. \ref{maspar}).

\begin{table}[htb]
\begin{center}
\renewcommand{\arraystretch}{1.5}
\begin{tabular}{|c|@{} p{0.1cm} @{}|c|c|c|c|}
\hline
 model && $\kappa$ & $\alpha$ & $\beta$ & $N_v$ \\
\hline
\hline
 CJK   && $1.934^{+0.112}_{-0.103}$  & $0.299^{+0.061}_{-0.051}$ & $0.898^{+0.204}_{-0.156}$ & $0.404^{+0.066}_{-0.054}$ \\
\hline
\end{tabular}
\caption{The parameters of the fits for CJK model with errors calculated in 
the Hessian quadratic approximation for the standard requirement of 
$\Delta \chi^2 = 1$}
\label{parerrors}
\end{center}
\end{table}

All test parton distributions along with \fun are further parametrized on 
the grid. The resulting FORTRAN program can be found on the web-page 
\cite{webpage}.


\subsection{Uncertainties of the CJK parton densities}

In this section we discuss the uncertainties of the CJK model parton 
densities.

\begin{figure}[h]
\includegraphics[scale=1.0]{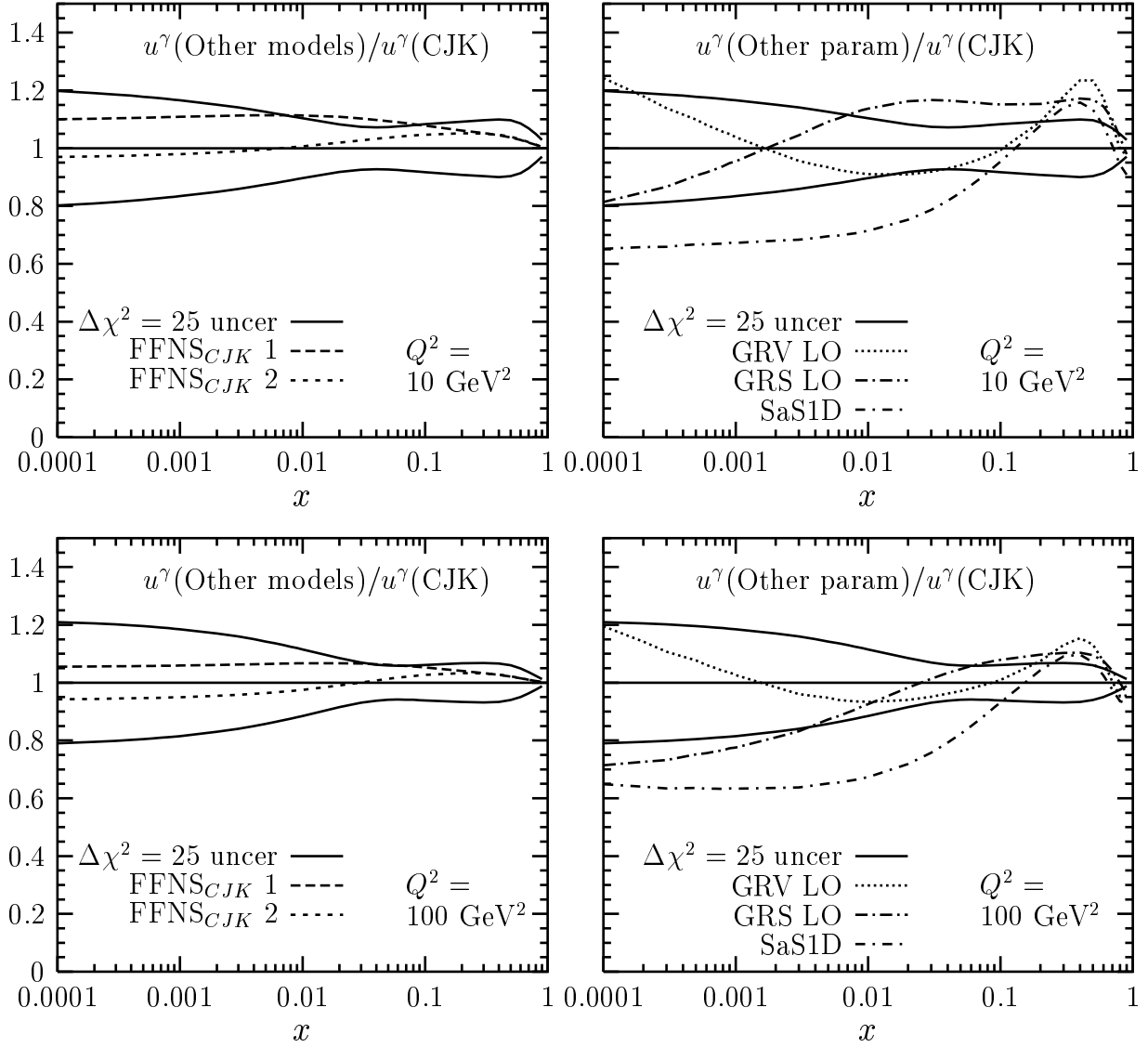}
\caption{Up-quark densities calculated in FFNS$_{CJK}$ models and
GRV LO \cite{grv92}, GRS LO \cite{grs} and SaS1D \cite{sas} parametrizations
compared with the CJK predictions. We plot for $Q^2=10$ and 100 GeV$^2$ the 
$u\gam(\mathrm{Other \: model})/u\gam(\mathrm{CJK})$ and
$u\gam(\mathrm{Other \: parametrization})/u\gam(\mathrm{CJK})$ ratios of the 
up-quark density calculated in the CJK model and its values obtained with 
other models and parametrizations. Solid lines show the CJK fit uncertainties 
for $\Delta \chi^2 = 25$ computed with the set of $\{S_i^{\pm}\}$ test
parametrizations.}
\label{densuncup}
\end{figure}

\begin{figure}[h]
\includegraphics[scale=1.0]{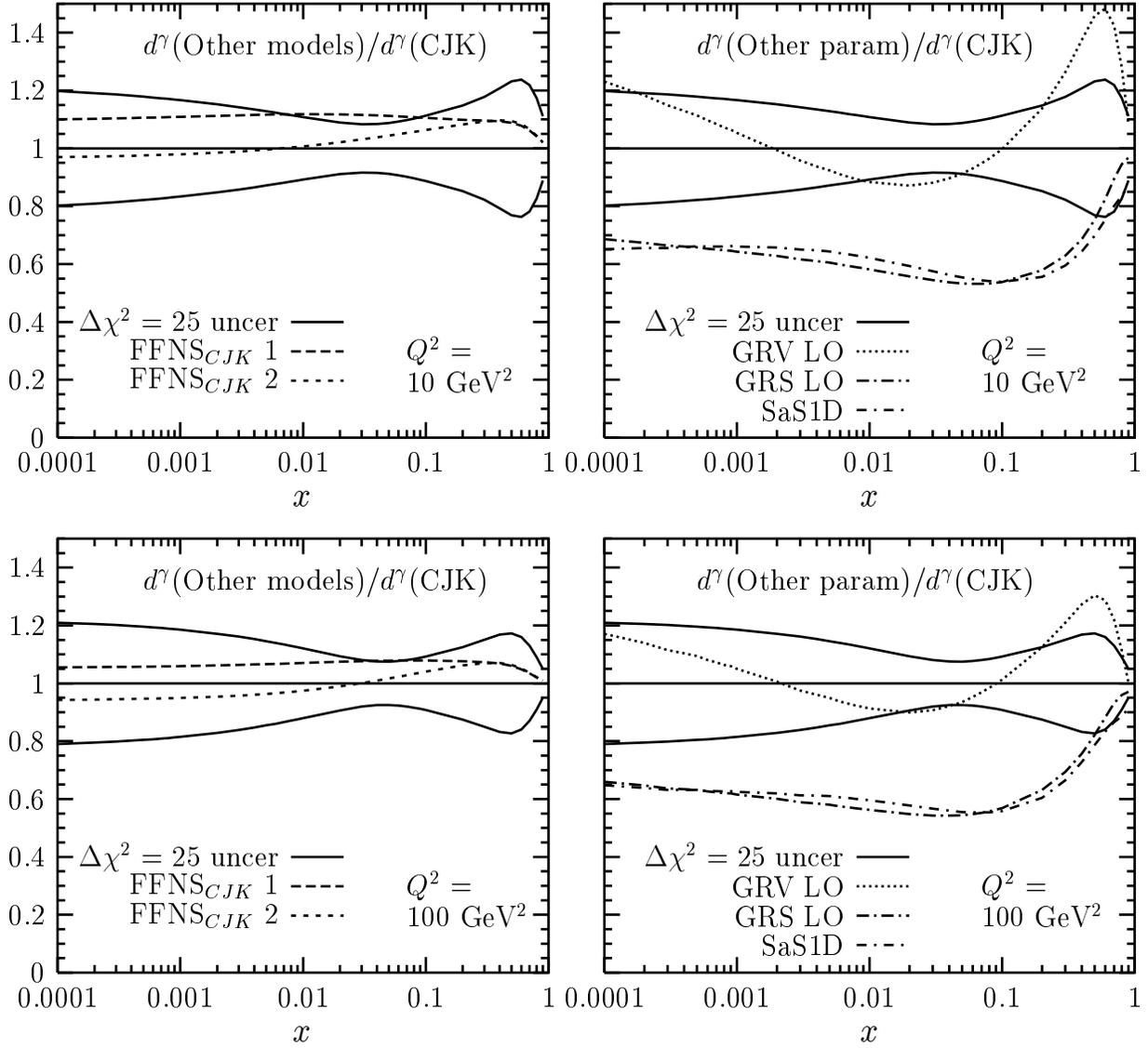}
\caption{The same as in Fig. \ref{densuncup}, for the down quark.}
\label{densundown}
\end{figure}

\begin{figure}[h]
\includegraphics[scale=1.0]{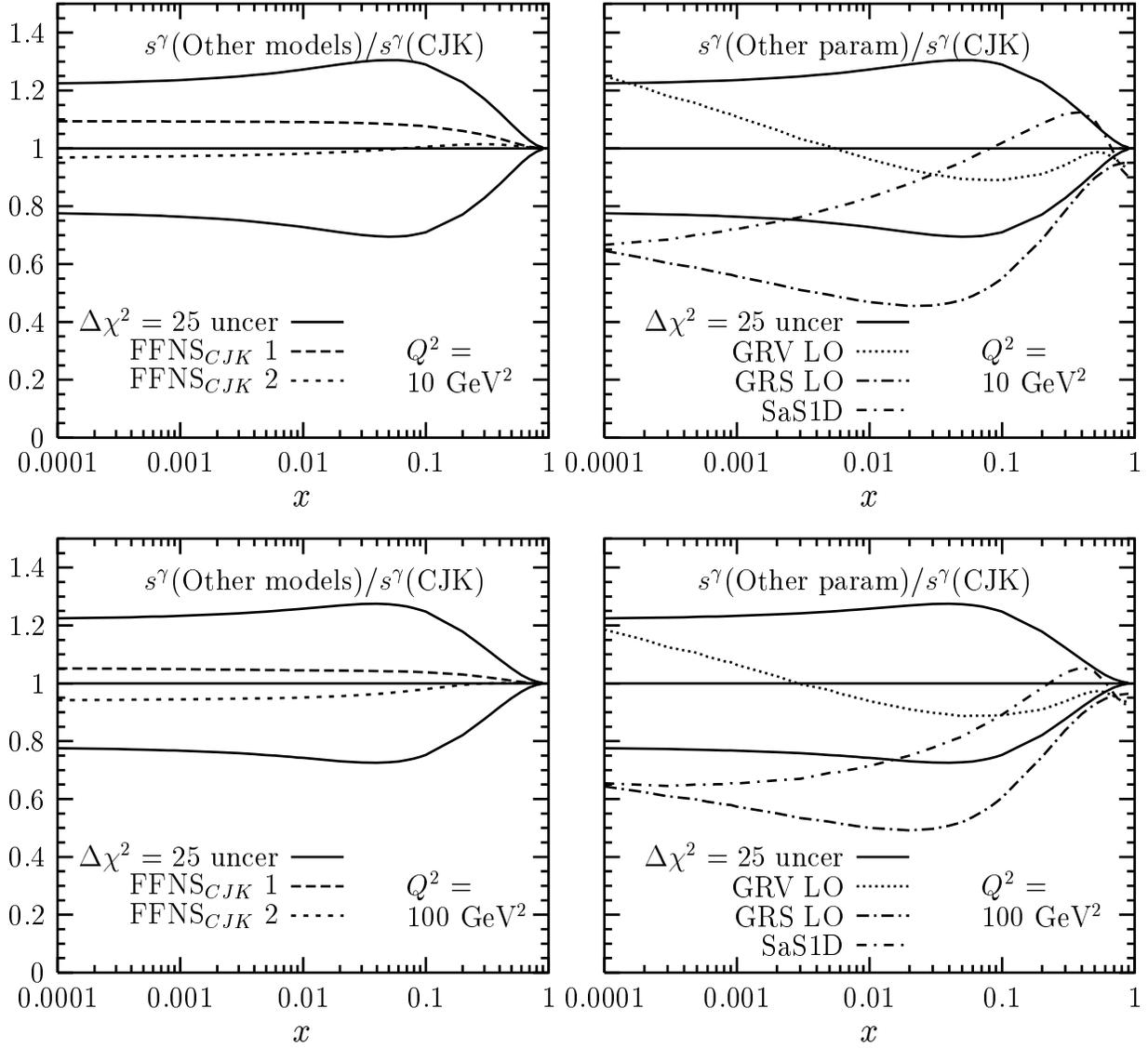}
\caption{The same as in Fig. \ref{densuncup}, for the strange quark.}
\label{densuncst}
\end{figure}

\begin{figure}[h]
\includegraphics[scale=1.0]{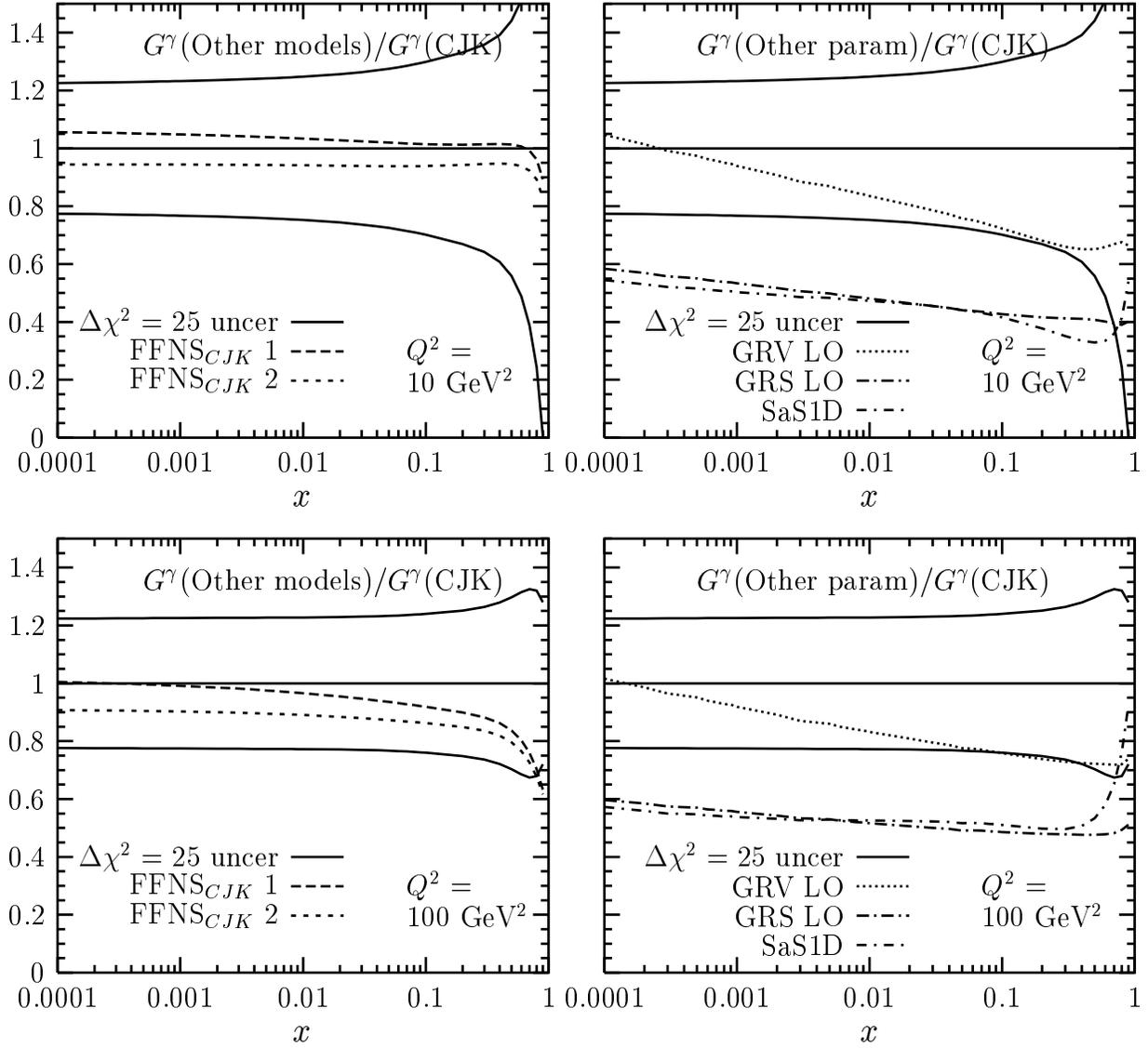}
\caption{The same as in Fig. \ref{densuncup}, for gluon.}
\label{densuncgl}
\end{figure}

\begin{figure}[h]
\includegraphics[scale=1.0]{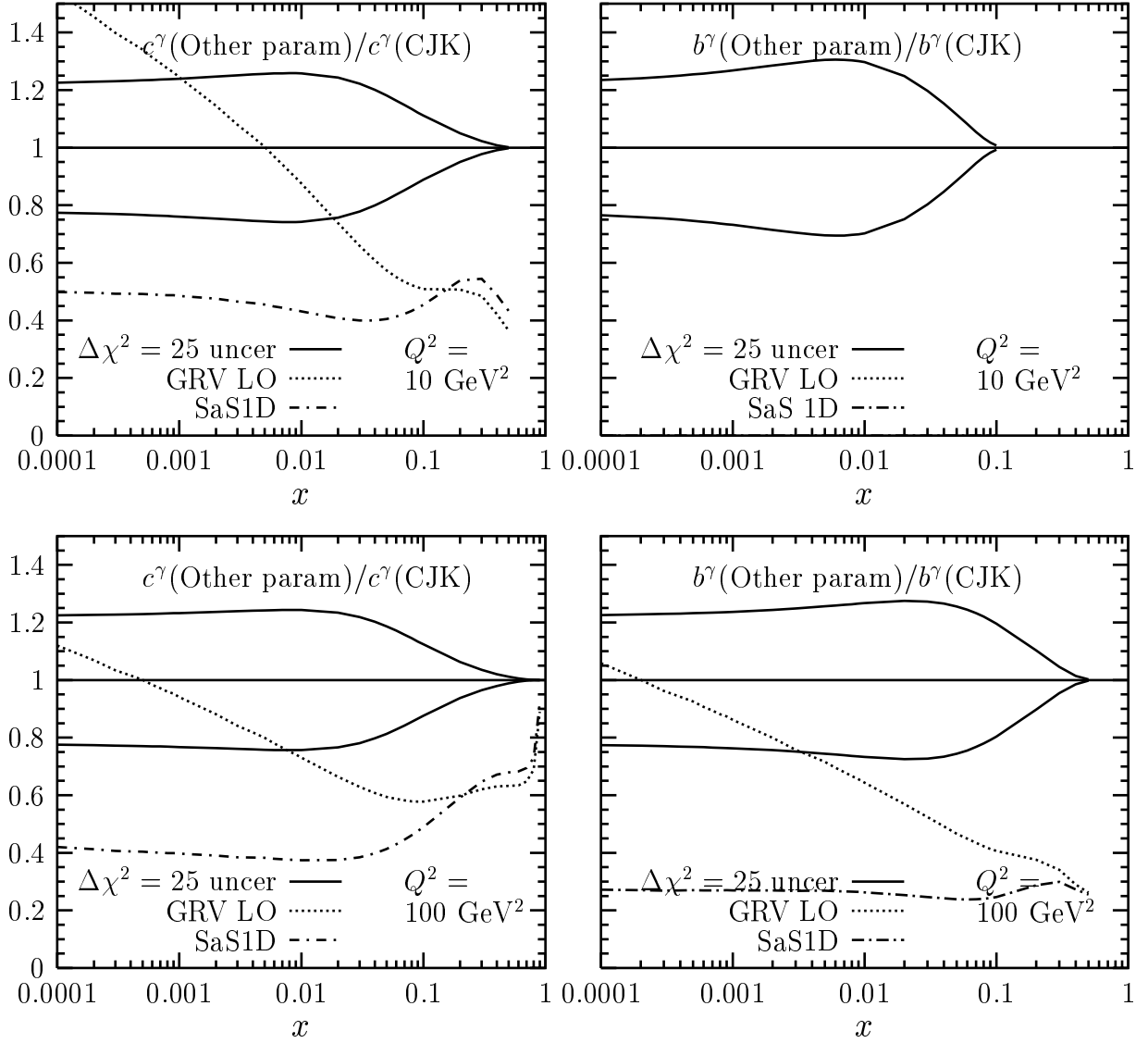}
\caption{The same as in Fig. \ref{densuncup}, for the charm and beauty quark. 
Only parametrizations including heavy-quark distributions are compared with 
the CJK model predictions.}
\label{densuncch}
\end{figure}

In Figures \ref{densuncup}--\ref{densuncch} the quark and gluon densities 
calculated in FFNS$_{CJK}$ models and GRV LO \cite{grv92}, GRS LO \cite{grs}
and SaS1D \cite{sas} parametrizations are compared with the CJK predictions. 
We plot for $Q^2=10$ and 100 GeV$^2$ the 
$q\gam(\mathrm{Other \: model})/q\gam(\mathrm{CJK})$ and\\
$q\gam(\mathrm{Other \: parametrization})/q\gam(\mathrm{CJK})$ ratios of 
the parton $q\gam$ densities calculated in the CJK model and their values 
obtained with other models and parametrizations. Solid lines show the CJK fit 
uncertainties for $\Delta \chi^2 = 25$ computed with the $\{S_i^{\pm}\}$ test
parametrizations.

First we notice that there is only one range of $x$, namely 
$0.01\simkl x \simkl 0.1$ at $Q^2=10$ GeV$^2$ where the up- and down-quark 
densities predicted by the FFNS$_{CJK}$ 1 fit go slightly beyond the 
uncertainty bands. Apart from that the predictions of FFNS$_{CJK}$ models in 
the case of all parton distributions lie between the lines of the CJK 
uncertainties. That indicates that the choice of $\Delta \chi^2 = 25$ agrees 
with the differences among our three models. Moreover the GRV LO 
parametrization predictions are nearly contained within the CJK model 
uncertainties. That is not the case only for the heavy-quark densities. To 
hold the GRS LO and SaS1D parametrization curves in that range would require 
a much increased $\Delta \chi^2$ value. Especially the SaS1D results differ 
very substantially from the CJK ones.

We observe that the up-quark distribution is the one best constrained by the
experimental data. As could be expected the greatest uncertainties are
connected with the gluon densities. In the case of $u\gam$ the 
$\Delta \chi^2 = 25$ band widens in the small $x$ region. Alike in the case of
other quark uncertainties it shrinks at high-$x$. On contrary the 
gluon distributions are least constrained at the region of $x\to 1$. That 
results from the fact that the gluon density contributes to \fun mainly through
the $\gamma^* G \to q\bar q$ process which gives numerically important
results only at small-x and therefore only in that region experimental data
can constrain it. Finally we see that all uncertainties become slightly 
smaller when we go to higher $Q^2$ from 10 to 100 GeV$^2$.


\section{Lagrange method for the uncertainties of the parton distributions}

The Hessian method, discussed above, is a very useful tool as it allows 
for computing of the parton density uncertainties in a very simple and 
effective way. If we find that our $\Delta \chi^2$ was assumed too rigorously 
or on contrary too conservatively we can obtain the uncertainties 
corresponding to any other $\Delta \chi^2$ value by simple scaling of the 
previous results. Still, the Hessian method relies on the assumption of 
the quadratic approximation, which in the case of our analysis is not 
perfectly preserved. Therefore, it is very important to perform a cross-check
of the obtained results by comparing them to the corresponding results derived
in a different statistical approach.

There exist another method called the Lagrange multiplier method (the 
Lagrange method in short) which allows to find exact uncertainties 
independently of the quadratic approximation. This method has also been 
applied to the proton structure case by the CTEQ Collaboration \cite{cteq1},
\cite{cteq4} and the MRST group \cite{mrst}. Here we utilize it to perform 
tests of the reliability of the Hessian approach results in the case of the 
CJK parton distributions.

In the Lagrange method we make a series of fits on the quantity
\be
F(\lambda,\{a_i\}) = \chi^2(\{a_i\}) + \lambda X(\{a_i\}),
\end{equation}
each with a different but fixed value of the Lagrange multiplier $\lambda$. As
a result we obtain a set of points $(\chi^2(\lambda),X(\lambda))$ which 
characterize the deviation of the physical quantity $X$ from its best value 
$X_0$ for a corresponding deviation of the structure function global fit from 
its minimum $\Delta \chi^2 = \chi^2(\lambda) - \chi_0$. In each of these 
constrained (by the $\lambda$ parameter) fits we find the best value of $X$ 
and the optimal $\chi^2$. For $\lambda = 0$ we return to the basic structure 
function fit which gave us $\{a_i^0\}$ parameters and allowed to calculate the
best $X_0$ value. The great advantage of this approach lies in the fact that 
we do not assume anything about the uncertainties. There is neither the 
quadratic nor any other approximation in that case. The large computer time 
consuming of the process of the whole series of minimalizations is its huge 
disadvantage.


\section{Examples of cross-section uncertainties in Hessian and Lagrange 
methods}

As was already said the Lagrange multiplier approach is a method on which we 
can rely whether the assumption of quadratic approximation of the Hessian is 
fulfilled or not. Unfortunately the amount of necessary computer calculations 
makes this approach very impractical. Still it can serve us as a final check 
of the correctness of the uncertainties that can be obtained using collections
of our $\{S_i^{\pm}\}$ parametrizations calculated with the Hessian method. 
We will show a comparison of the uncertainties obtained in Hessian and 
Lagrange multiplier methods calculated with the CJK parton distributions for 
two physical quantities. First, for the one of $F\gam_{2,c}$ points
measured by the OPAL Collaboration \cite{F2c}, as described in Section 4.3. 
Secondly, for the $\gamma q \to \gamma q$ part of the $\gamma \gamma$ prompt 
photon production cross-section.


\subsection{$F\gam_{2,c}$}

We chose to exam our collection of test parametrizations first on a very simple
example of $F\gam_{2,c}$. The charm-quark structure function depends only on 
the charm-quark and gluon distributions which (especially the gluon density) 
are not well constrained by the experimental data. On the other hand this 
dependence is not very strong in the high-$x$ region where the direct 
Bethe-Heitler contribution dominates. Still we expect considerable deviation 
of the Hessian method results from the correct Lagrange approach predictions.

\begin{figure}[h]
\begin{center}
\input{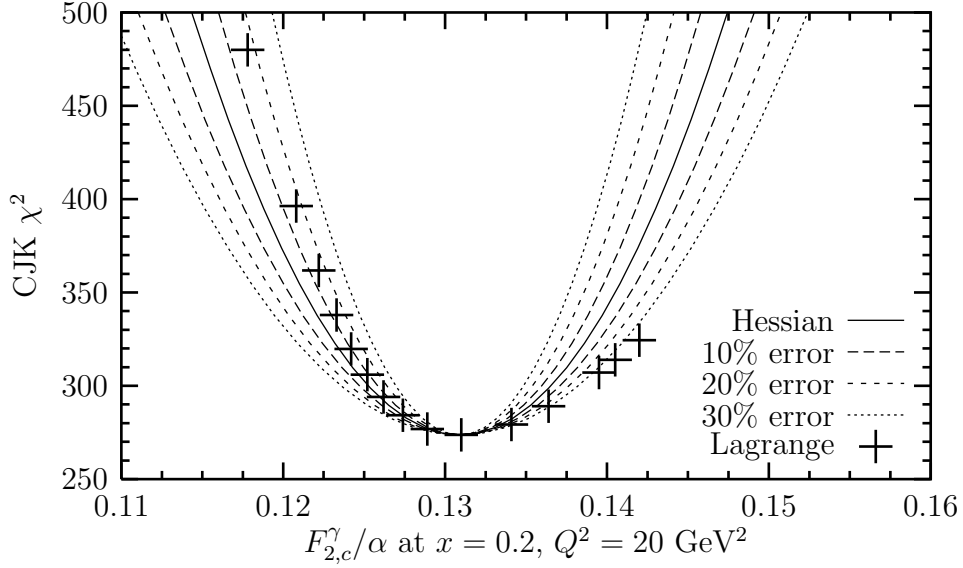} 
\caption{The comparison of the Lagrange and Hessian method results for
the high bin $F\gam_{2,c}$ of the OPAL \cite{F2c} meauserement. The solid line
and crosses show the Hessian and Lagrange method predictions respectively. The
dashed lines represent the 10 to 30\% deviation from the Hessian result.}
\label{f2c1}
\end{center}
\end{figure}

The \cite{F2c} OPAL analysis provided the averaged $F\gam_{2,c}/\alpha$ values
in two $x$ bins. For our purpose we chose the high-$x$ point at $x=0.2$ and
$Q^2=20$ GeV$^2$. We calculate the CJK model prediction for 
$F\gam_{2,c}/\alpha$ and its Hessian uncertainties. Further we apply the
Lagrange method with $\lambda = \pm 0.33,0.66,1,\cdots \times 10000$. The
results are presented in Fig. \ref{f2c1}. The solid line and crosses show the 
Hessian and Lagrange method predictions respectively. The dashed lines 
represent the 10 to 30\% deviation from the Hessian results. As can be seen 
the Hessian method reproduces very well the Lagrange predictions in the 
direction of decreasing $F\gam_{2,c}$ even for $\Delta \chi^2$ greater than 
100. On the other hand the agreement in the other direction corresponding to 
negative $\lambda$ is much worse. The lack of higher $\chi^2$ Lagrange 
points results from the negative $N_v$ parameter values appearing in the fits 
for $\lambda > 1 \times 10000$.


\subsection{Prompt photon production}

As the second example physical process we took a part of the prompt photon 
$\gamma \gamma$ production. Namely, for the sake of the limitation of the
necessary computer time, we decided to calculate the direct resolved (DR)
$\gamma q \to \gamma q$ cross-section. Any cross-section of two photons
can be divided into three parts. The direct part which is a simple 
$\gamma \gamma$ electromagnetic interaction, the DR being the part in which 
one photon interacts with the partons originating from the second $\gamma$
and finally the doubly resolved (or resolved resolved, RR) part in which
both photons interact through their hadronic structure. Only the resolved
processes contribute to the prompt photon production. Unfortunately calculation
of the RR part is very computer time consuming. Therefore as we are not 
interested in the quantitative result of our computation but we use it only to
compare the uncertainties obtained in two methods we omit the doubly resolved 
contribution to the cross-section. The $\gamma q \to \gamma q$ process 
depends only on the quark distributions. We include the heavy-quark
contributions with omitting their masses in calculations (the difference to 
the massive computation is on a 1\% level). We calculate the cross-section for
the photon beams energy $E_{\gamma}=200$ GeV. That can correspond to the high 
energy peak of the TESLA Photon Collider \cite{tesla} built on the Linear 
Collider of the $e^+e^-$ central mass energy of 500 GeV.

\begin{figure}[h]
\begin{center}
\input{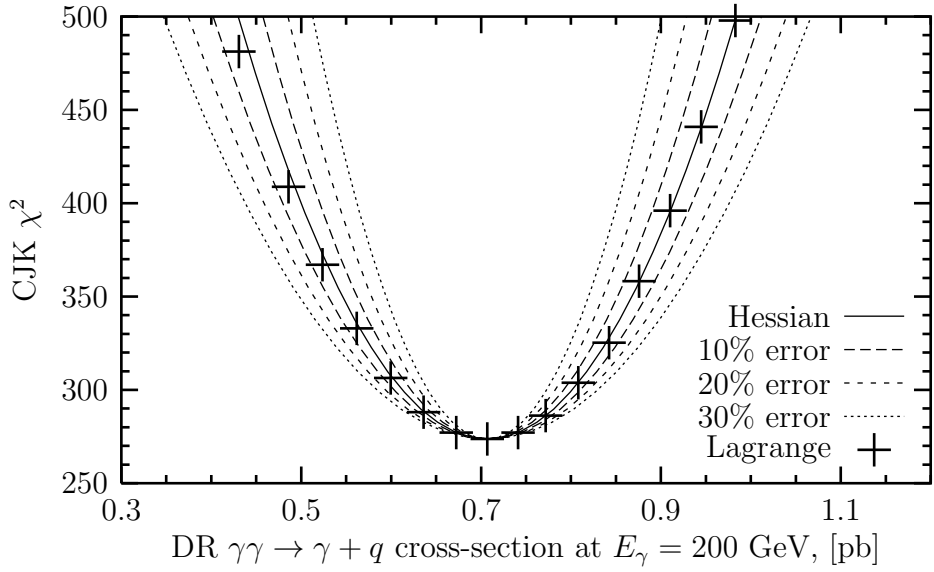} 
\caption{The comparison of the Lagrange and Hessian method results for
the direct resolved (DR) part of the $\gamma \gamma \to \gamma q$ 
cross-section. The solid line and crosses show the Hessian and Lagrange method
predictions respectively. The dashed lines represent the 10 to 30\% deviation 
from the Hessian result.}
\label{prompt1}
\end{center}
\end{figure}

The Lagrange method was applied with 
$\lambda = \pm 0.2,0.4,0.6,\cdots \times 1000$. The results of the comparison 
are presented in Fig. \ref{prompt1}. Again the solid line and crosses show the
Hessian and Lagrange method predictions respectively. The dashed lines 
represent the 10 to 30\% deviation from the Hessian results. As we observe 
both methods agree very well in both directions of the change of the 
cross-section and in the whole range of the $\Delta \chi^2$ plotted.


\section{$\chi^2$ of the CJK fit and the data \label{sectdata}}

The CJK model presented in this article is, as described in detail in 
\cite{cjk}, a result of an improved analysis of \cite{cjkl} where our first
CJKL model was given. Though we obtain a much better $\chi^2/_{\rm DOF}$
in the case of the CJK global fit to \fun comparing to the previous CJKL fit,
its value, 1.537, is large. Moreover, part of the improvement of the value of 
$\chi^2/_{\rm DOF}$ is due to the exclusion of the TPC2$_{\gamma}$ data from 
the full set of the experimental $F_2\gam(x,Q^2)$ results. Recently a similar 
global fit has been performed and presented in \cite{klasen} for the case of 
the simple FFNS type model with the Bethe-Heitler process describing the 
heavy-quark contributions. The result of that fit is $\chi^2/_{\rm DOF}=0.938$.
The authors applied 134 \fun points. In this section we try to indicate the 
main sources of the large $\chi^2/_{\rm DOF}$ of our CJK fit to 182 data 
points. We also try to judge whether any exclusions of the data from the fits 
can be justified.

We performed a few simple tests. First, for the case of the additional 
test global fit containing the full set of the available data, mentioned in 
section \ref{secttolpar} and denoted further as FULL. We calculated the 
$\chi^2_n/N_n$ for each of the data collections as explained in 
\ref{secttolpar}. The $\chi^2_n$, being the value of the $\chi^2$ of the best 
fit corresponding to the given experiment, divided by the number of the 
experimental points $N_n$ shows the agreement of each of the data collections 
with the global fit. Results are presented in the table \ref{tabdata}. 
Obviously, as different sets of the data were applied in this and CJK
fits the corresponding numbers in tables \ref{chi2t} and \ref{tabdata} are not
the same. The FULL fit results show that the global test similarly disagree 
with three of the data sets, the TPC2$_{\gamma}$, DESY and DELPHI.

\begin{table}[htb]
\begin{center}
\renewcommand{\arraystretch}{1.5}
\begin{tabular}{|c|@{} p{0.1cm} @{}|c|c|c|c|c|c|c|}
\hline
 Set          && TPC2$_{\gamma}$ & DESY & KEK & ALEPH & DELPHI & L3 & OPAL \\
\hline
\hline
 $\chi^2_n/N_n$ && 2.38 & 2.44 & 1.20 & 1.07 & 2.25 & 0.60 & 0.97 \\
\hline
\end{tabular}
\caption{The $\chi^2_n$, being the value of the $\chi^2$ of the best fit 
corresponding to the given experiment, divided by the number of 
the experimental points $N_n$ for the case of the test global FULL fit to the 
full set of the available data.}
\label{tabdata}
\end{center}
\end{table}

Secondly, we calculated the contribution to the $\chi^2$ of the FULL fit given
by each of the 208 data points. We performed the same computation for the GRS 
\cite{grs} and SaS1D \cite{sas} parametrizations as well as for the two 
FFNS$_{CJK}$ fits presented in \cite{cjk} (obtained without the 
TPC2$_{\gamma}$ data). We found the data which contribution to the $\chi^2$ in
case of each of the mentioned fits is larger than 3. It appeared that there 
are 5 such points among 12 in the case of the CELLO experiment, \cite{CELLO}, 
belonging to the DESY collection. Moreover, there are 9 such points among 22 
in the case of the DELPHI'01 experiment, \cite{DELPHI01}, belonging to the 
DELPHI collection. On the other hand we found only 2 such points in the case 
of the TPC2$_{\gamma}$ data. In other experiments there are no more than 
single cases of that kind.

As the DESY and DELPHI sets are the collections of various measurements
(in the DELPHI experiment case they differ by the year of publication) we
checked in more detail the $\chi^2_n/N_n$ for each of those measurements in the
case of the FULL fit. The results, presented in the table \ref{desdel}, 
confirm that as could be expected from the above test the CELLO and DELPHI'01 
results give main contributions to the high $\chi^2_n/N_n$ of the DESY and 
DELPHI collections respectively.

\begin{table}[htb]
\begin{center}
\renewcommand{\arraystretch}{1.5}
\begin{tabular}{|c|@{} p{0.1cm} @{}|c|c|c|c|c|c|c|}
\hline
 Set          && PLUTO & JADE & CELLO & TASSO & DELPHI'96 & DELPHI'98 & DELPHI'01 \\
\hline
\hline
 $\chi^2_n/N_n$ && 0.64 & 1.12 & 5.88 & 1.01 & 1.04 & 0.10 & 3.65 \\
\hline
\end{tabular}
\caption{The $\chi^2_n$, being the value of the $\chi^2$ of the best fit 
corresponding to the given experiment, divided by the number of 
the experimental points $N_n$ for the case of the global FULL fit to the full 
set of the available data.}
\label{desdel}
\end{center}
\end{table}

\begin{figure}[h]
\includegraphics[scale=0.8]{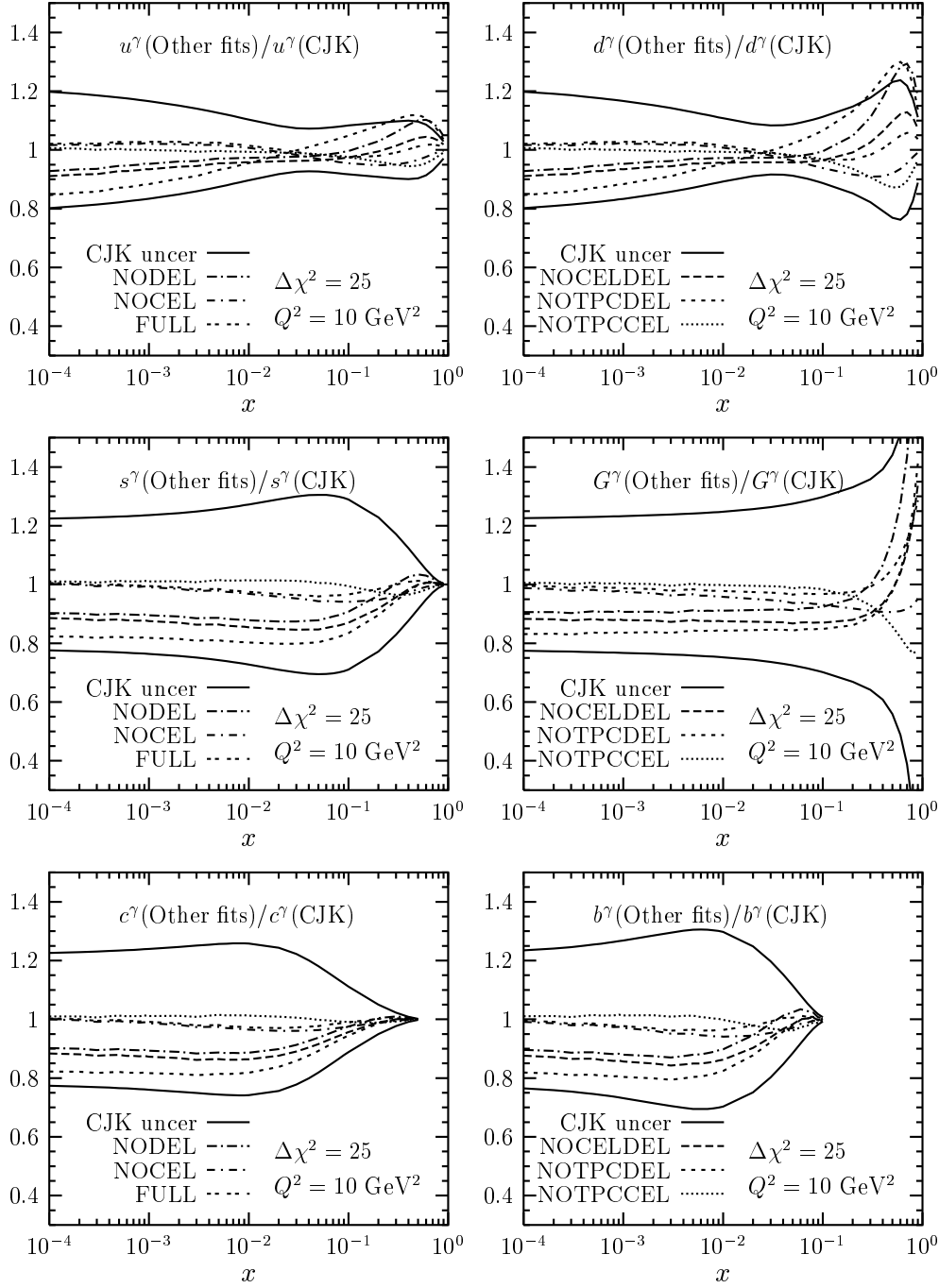}
\caption{Comparison of the test fit parton distributions with the CJK parton 
densities and their uncertainties. Fit denoted as FULL contains all the 
available data. The NOCEL, NODEL and CJK fits exclude the CELLO \cite{CELLO}, 
DELPHI'01 \cite{DELPHI01} and TPC2$_{\gamma}$ \cite{TPC} measurements 
respectively. The NOTPCCEL, NOTPCDEL and NOCELDEL fits exclude pairs of CELLO,
DELPHI'01 and TPC2$_{\gamma}$ data. We plot for $Q^2=10$ GeV$^2$ the 
$q\gam(\mathrm{Other \: fits})/q\gam(\mathrm{CJK})$ ratios of densities 
calculated in the CJK fit and their values obtained in other fits. Solid lines
show the CJK fit uncertainties for $\Delta \chi^2 = 25$ computed with the set 
of $\{S_i^{\pm}\}$ test parametrizations.}
\label{figdata}
\end{figure}

Further, we performed another five test fits of the CJK model to the 
full data set from which we excluded the CELLO, DELPHI'01 and finally pairs
of CELLO, DELPHI'01 and TPC2$_{\gamma}$ measurements. We denote those fits as 
NOCEL, NODEL (CJK fit = NOTPC) and NOTPCCEL, NOTPCDEL, NOCELDEL respectively. 
The $\chi^2$, the number of experimental points and $\chi^2/_{\rm DOF}$ for 
the CJK and each of the six test fits are presented in table \ref{datachi2}. 
The FULL fit including all available data has highest $\chi^2/_{\rm DOF}$
value. The exclusion of the TPC2$_{\gamma}$ measurement, proposed 
in \cite{klasen}, improves it but not very much, see the CJK fit result.
The NOCEL and NODEL fits give much better agreement between the model and the 
data but still in their case $\chi^2/_{\rm DOF}$ is larger than 1. We see 
that only exclusion of two of the CELLO, DELPHI'01 and TPC2$_{\gamma}$ 
measurements at the same time leads to the $\chi^2/_{\rm DOF} \approx 1$. 
Finally, we notice that the best $\chi^2/_{\rm DOF}$ is obtained in the case 
of the NOCELDEL fit. Central values of the parameters calculated in the 
additional test fits (not shown) and of the CJK fit (given in table 
\ref{tparam}) lie inside the sum of their uncertainties. This statement is 
true for both cases of the CJK fit uncertainties: obtained from \textsc{Minos}
(see table \ref{tparam}) and calculated in the Hessian quadratic approximation
(see table \ref{parerrors}). Only differences between the central values of 
the $\beta$ parameter are slightly larger than the sums of the corresponding 
uncertainties. Therefore, we examined the agreement among the parton 
distributions computed in the CJK model in each of the fits. Figure 
\ref{figdata} presents the comparison of the six test fit results with the 
CJK parton densities and their uncertainties obtained in the 
Hessian method for $\Delta \chi^2 = 25$ and $Q^2=10$ GeV$^2$. We notice that 
the test distributions are in agreement with the CJK densities in the sense 
that they are contained within the CJK uncertainties for the assumed allowed 
deviation of the global fit from the minimum. Especially, we observe a very 
small difference among the CJK, FULL, NOCEL and NOTPCCEL parton distributions.
The only deviation is observed in the case of the up- and down-quark densities
at high-$x$ computed for the NODEL and NOTPCDEL fits. The results at other 
$Q^2$ values are very similar.

\begin{table}[htb]
\begin{center}
\renewcommand{\arraystretch}{1.5}
\begin{tabular}{|c|@{} p{0.1cm} @{}|c|@{} p{0.1cm} @{}|c|c|c|@{} p{0.1cm} @{}|c|c|c|}
\hline
 FIT                 && FULL  && CJK  & NOCEL & NODEL && NOTPCCEL & NOTPCDEL & NOCELDEL \\
\hline
\hline
 $\chi^2$            && 338.7 && 273.7 & 262.4 & 244.0 && 200.7 & 169.4 & 179.5 \\
\hline
points   && 208   && 182   & 196   & 186   && 170   & 160   & 174   \\
\hline
 $\chi^2/_{\rm DOF}$ && 1.66  && 1.54  & 1.37  & 1.34  && 1.21  & 1.08  & 1.06  \\
\hline
\end{tabular}
\caption{The $\chi^2$, the number of experimental points and 
$\chi^2/_{\rm DOF}$ for CJK fit and six additional test fits. Fit denoted as 
FULL contains all the available data. The NOCEL, NODEL and CJK fits exclude 
the CELLO \cite{CELLO}, DELPHI'01 \cite{DELPHI01} and TPC2$_{\gamma}$ 
\cite{TPC} measurements respectively. The NOTPCCEL, NOTPCDEL and NOCELDEL fits
exclude pairs of CELLO, DELPHI'01 and TPC2$_{\gamma}$ data.}
\label{datachi2}
\end{center}
\end{table}

We notice that the value of $\chi^2/_{\rm DOF}$ for the global 
fits of our CJK model depends very strongly on the choice of the data set 
applied in the fit, on the other hand, they produce similar parton 
distributions which lie within the CJK parton density uncertainties. 
Especially, as was noticed, the parton distributions predicted by our CJK and 
by the FULL fit, including all available \fun data, differ very slightly.


\section{Summary}

The very first analysis of the uncertainties of the radiatively generated 
parton distributions in the real photon based on the LO DGLAP equations is 
given. We consider the CJK model presented in \cite{cjk}. The estimate of the 
uncertainties of the CJK parton densities due to the experimental errors is 
based on the Hessian method which was recently applied in the proton parton 
structure analysis. We test the applicability of the approach by comparing 
uncertainties of example cross-sections calculated in the Hessian and Lagrange 
methods. Sets of test parametrizations are given, which allow for calculation 
of its best fit parton distributions along with \fun and for computation of 
uncertainties of any physical value depending on the real photon parton 
densities. Finally, we present a detailed analysis of the $\chi^2$ of the CJK 
fit and its relation to the data. We show that large $\chi^2/_{\rm DOF}$ of 
the fit is due to only a few of the experimental measurements. By excluding 
them $\chi^2/_{\rm DOF}\approx 1$ can be obtained. A FORTRAN program with the 
grid parametrization of the test parton distributions and \fun can be obtained 
from the web-page \cite{webpage}.

Our work is the first trial to estimate the uncertainties of the real photon 
parton distributions. Future analysis of that kind should include the 
corrections taking into account all possible sources of the deviation of the 
experimental \fun data from the photon structure function as described by the 
theoretical models.


\section*{Acknowledgments}

P.J. would like to thank M.Krawczyk, F.Cornet and R.Nisius for discussions and
important comments and remarks.

Moreover P.J. is thankful to J.Jankowska and M.Jankowski for their remarks on 
the numerical method applied in the grid parametrization program and 
A.Zembrzuski for further useful suggestions.



\clearpage

\clearpage

\clearpage

\clearpage

\clearpage

\clearpage

\clearpage

\clearpage

\end{document}